\documentclass[iop,apj]{emulateapj}

\usepackage{acronym}
\usepackage{booktabs}
\usepackage{dcolumn}
\usepackage{graphicx}
\usepackage{amsmath,amsfonts,amssymb}
\usepackage{mathrsfs}
\usepackage[utf8]{inputenc}
\usepackage[normalem]{ulem}

\usepackage[breaklinks,colorlinks,citecolor=blue]{hyperref}

\newcolumntype{d}[1]{D{.}{.}{#1}}

\usepackage{natbib}
\newcommand{\cf}{\textit{cf.}}
\newcommand{\ie}{\textit{i.e.}}
\newcommand{\eg}{\textit{e.g.}}

\usepackage{color}

\newacro{ADM}{Arnowitt-Deser-Misner}
\newacro{AMR}{adaptive mesh-refinement}
\newacro{BH}{black hole}
\newacro{BBH}{binary black-hole}
\newacro{BHNS}{black-hole neutron-star}
\newacro{BNS}{binary neutron star}
\newacro{CCSN}{core-collapse supernova}
\newacroplural{CCSN}[CCSNe]{core-collapse supernovae}
\newacro{CMA}{consistent multi-fluid advection}
\newacro{DG}{discontinuous Galerkin}
\newacro{HMNS}{hypermassive neutron star}
\newacro{EM}{electromagnetic}
\newacro{ET}{Einstein Telescope}
\newacro{EOB}{effective-one-body}
\newacro{EOS}{equation of state}
\newacroplural{EOS}[EOSs]{equations of state}
\newacro{FF}{fitting factor}
\newacro{GR}{general relativity}
\newacro{GRLES}{general-relativistic large-eddy simulation}
\newacro{GRHD}{general-relativistic hydrodynamics}
\newacro{MHD}{magnetohydrodynamics}
\newacro{GRMHD}{general-relativistic magnetohydrodynamics}
\newacro{GW}{gravitational wave}
\newacro{kN}{kilonova}
\newacroplural{kN}[kNe]{kilonovae}
\newacro{ILES}{implicit large-eddy simulations}
\newacro{ISM}{interstellar medium}
\newacro{LIA}{linear interaction analysis}
\newacro{LES}{large-eddy simulation}
\newacroplural{LES}[LES]{large-eddy simulations}
\newacro{MRI}{magnetorotational instability}
\newacro{NIR}{near infrared}
\newacro{NR}{numerical relativity}
\newacro{NS}{neutron star}
\newacro{PNS}{protoneutron star}
\newacro{SASI}{standing accretion shock instability}
\newacro{SGRB}{short $\gamma$-ray burst}
\newacro{SN}{supernova}
\newacroplural{SN}[SNe]{supernovae}
\newacro{SNR}{signal-to-noise ratio}

\begin{document}

\title{Viscous-Dynamical Ejecta from Binary Neutron Star Mergers}

\author{David Radice\altaffilmark{1,2},
Albino Perego\altaffilmark{3,4},
Kenta Hotokezaka\altaffilmark{2},
Sebastiano Bernuzzi\altaffilmark{5,3},\\
Steven A. Fromm\altaffilmark{6}
and Luke F. Roberts\altaffilmark{6}}
\altaffiltext{1}{Institute for Advanced Study, 1 Einstein Drive,
Princeton, NJ 08540, USA}
\altaffiltext{2}{Department of Astrophysical Sciences, Princeton University,
4 Ivy Lane, Princeton, NJ 08544, USA}
\altaffiltext{3}{Istituto Nazionale di Fisica Nucleare, Sezione Milano
Bicocca, gruppo collegato di Parma, I-43124 Parma, Italy}
\altaffiltext{4}{Dipartimento di Fisica, Universit\`{a} degli Studi di
Milano Bicocca, Piazza della Scienza 3, 20126 Milano, Italia}
\altaffiltext{5}{Theoretisch-Physikalisches
  Institut,Friedrich-Schiller-Universit{\"a}t Jena, 07743, Jena,
  Germany}
\altaffiltext{6}{NSCL/FRIB and Department of Physics \& Astronomy,
  Michigan State University, 640 S Shaw Lane East Lansing, MI 48824,
  USA}

\begin{abstract}
General-relativistic simulations of binary neutron star mergers with
viscosity reveal a new outflow mechanism operating in unequal mass
binaries on dynamical timescales and enabled by turbulent viscosity.
These ``viscous-dynamical'' ejecta are launched during merger due to the
thermalization of mass exchange streams between the secondary and the
primary neutron star.
They are characterized by asymptotic velocities extending up to ${\sim}
0.8\, c$, and have masses that depend on the efficiency of the
viscous mechanism.
Depending on the unknown strength of the effective viscosity arising from
magnetohydrodynamics instabilities operating during merger, the overall
mass of the dynamical ejecta could be enhanced by a factor of
a few and the mass of the fast tail of the ejecta having asymptotic
velocities $\geq 0.6\, c$ by up to four orders of magnitude.
The radioactive decay of the expanding viscous-dynamical ejecta could
produce bright kilonova transients with signatures of free neutron decay
in the first hour and enhanced near infrared flux on a timescale of
a few days.
The synchrotron remnant produced by the interaction between the ejecta
and the interstellar medium could also be significantly enhanced by
viscosity. Such remnant could be detected in the case of GW170817 as a
rebrightening of the radio signal in the next months to years.
\end{abstract}


\section{Introduction}

The UV/optical/infrared counterpart \citep{chornock:2017sdf,
cowperthwaite:2017dyu, drout:2017ijr, nicholl:2017ahq, tanaka:2017qxj,
tanvir:2017pws, perego:2017wtu, villar:2017wcc, waxman:2017sqv} of the
binary \ac{NS} merger event GW170817 \citep{theligoscientific:2017qsa,
abbott:2018wiz, gbm:2017lvd} is thought to have been powered by the
radioactive decay of about $0.05\ M_\odot$ of material ejected during
and shortly after the merger \citep{lattimer:1974a, symbalisty:1982a,
meyer:1989a, eichler:1989ve, freiburghaus:1999a, goriely:2011vg,
korobkin:2012uy, wanajo:2014wha, just:2014fka, thielemann:2017acv,
kasen:2017sxr, rosswog:2017sdn, hotokezaka:2018aui}, the so called
\ac{kN}.

The observations can be fitted with a minimal two-components \ac{kN}
model (\citealt{chornock:2017sdf, cowperthwaite:2017dyu, drout:2017ijr,
nicholl:2017ahq, tanaka:2017qxj, tanvir:2017pws, villar:2017wcc}; see
however \citealt{waxman:2017sqv} for an alternative model). The first
component, often called the ``blue \ac{kN}'', peaked on a short
timescale (${\sim}1$~day from the merger) in the UV/optical bands and is
thought to have been powered by the radioactive decay of ${\sim}0.02\,
M_\odot$ of low photon-opacity material with a large expansion velocity
${\sim} 0.3\, c$. The second component, typically referred to as the
``red \ac{kN}'', peaked on a timescale of ${\sim}5$~days in the \ac{NIR}
bands and is thought to have been powered by the radioactive decay of
${\sim}0.04\, M_\odot$ of high photon-opacity material characterized by
a lower expansion velocity ${\sim} 0.1\, c$. This minimal model is,
however, incompatible with \ac{GRHD} merger simulations that cannot
produce a sufficient amount of fast moving ejecta
\citep[\eg,][]{davies:1993zn, rosswog:2012wb, hotokezaka:2012ze,
bauswein:2013yna, radice:2016dwd}. This discrepancy could point to the
presence of additional physics beyond that included in the simulations,
such as the presence of magnetized winds from the merger remnant
\citep{metzger:2018uni, fernandez:2018kax}. \ac{kN} models including
anisotropy, multiple ejecta components and their interactions can only
partially reconcile observations and simulations
\citep[\eg,][]{perego:2017wtu, kawaguchi:2018ptg}.

Merger simulations showed that binary \ac{NS} mergers generate outflows
through a number of different mechanisms operating on different
timescales. The dynamical ejecta are launched during merger by tidal
interactions and shocks exerted on the \acp{NS} on a dynamical timescale
\citep[\eg,][]{rosswog:1998hy, rosswog:2001fh, hotokezaka:2012ze,
bauswein:2013yna, wanajo:2014wha, sekiguchi:2015dma, radice:2016dwd,
sekiguchi:2016bjd}. They have masses of ${\sim} 10^{-4}{-}10^{-2}$
$M_\odot$ and typical velocities distributed in ${\sim} 0.1{-}0.3$~c
with ${\sim}10^{-6}{-}10^{-5}\, M_\odot$ of fast ejecta having
velocities larger than $0.6$~c. The superposition of dynamical ejecta
launched by different mechanisms results in an outflow with a wide range
of electron fractions $0.05<Y_e<0.4$, with higher $Y_e$ material being
typically channeled in the polar directions \citep[\cf~detailed
discussion in][]{radice:2016dwd, radice:2018pdn}.

Neutrino re-absorption by material ablated from the surface of the
\ac{HMNS} and/or from the accretion disk originates another outflow
component channeled along the polar direction $\theta \lesssim 45^\circ$
with $Y_e > 0.25$ \citep{dessart:2008zd, perego:2014fma, just:2014fka,
martin:2015hxa, perego:2017wtu}. More material is also expected to be
unbound by viscous and nuclear processes in the remnant's accretion disk
\citep{metzger:2008av, metzger:2008jt, lee:2009a, fernandez:2013tya,
siegel:2014ita, metzger:2014ila, martin:2015hxa, wu:2016pnw,
siegel:2017nub, lippuner:2017bfm, fujibayashi:2017xsz,
fujibayashi:2017puw, siegel:2017jug, metzger:2018uni, fernandez:2018kax,
radice:2018xqa}.  These additional secular ejecta are expected to
provide a dominant contribution to the \ac{kN} on timescales of days to
weeks, while the dynamical ejecta discussed above mostly contributes to
the early ${\sim}1$~day emission \citep{radice:2018pdn}. The presence
of a fast dynamical ejecta component with velocities in excess $0.6$~c
might be responsible for a brighter UV/Optical polar emission and a
precursors powered by the beta-decay of the free neutrons
\citep{metzger:2014yda}. The dynamical ejecta also generate synchrotron
radiation by the shock interaction with the interstellar medium
\citep[\eg,][]{nakar:2011cw, hotokezaka:2015eja, hotokezaka:2018gmo};
such radio signature depends on the medium density and on the kinetic
energy of the outflow. Current merger simulations cannot
self-consistently predict the properties of the secular ejecta,
especially because the treatment of key physical processes, such as
neutrino-matter interaction and angular momentum transport due to
magnetic effects, is limited by the use of approximate schemes and
insufficient resolution \citep[\eg,][]{foucart:2016rxm, kiuchi:2017zzg}.
The uncertainties in the theoretical modeling are crucially reflected in
the current challenges in the interpretation of the observations.

In this \emph{Letter} we report the finding of a new outflow mechanism
that can operate in unequal mass binaries on dynamical timescales and is
enabled by turbulent viscosity. We show that this mechanism can boost
the mass of the dynamical ejecta by a factor of a few. The resulting
viscous-dynamical ejecta are characterized by their larger than typical
mass and their distribution extending to high velocities. These outflows
could produce UV/optical transients on a timescale of a few hours from
the merger, in part powered by the decay of free neutrons in the
high-velocity tail of the ejecta. The viscous-dynamical ejecta would
also contribute to the overall \ac{NIR} flux of the \ac{kN}, and would
produce bright radio flares on timescales of weeks to years from the
merger. The magnitude of this outflow component for GW170817 could be
constrained by future radio observations in the near future.

\section{Method}

We study the role of the turbulent viscosity in the dynamical ejection
of mass during \ac{NS} mergers using the general-relativistic large eddy
simulations method (GRLES; \citealt{radice:2017zta}). Viscosity effects
are modeled by augmenting the perfect fluid stress-energy tensor with a
purely spatial tensor representing the effect of subgrid-scale
turbulence. The latter is effectively parametrized by the turbulent
viscosity coefficient $\nu_{_T} = \ell_{\rm mix} c_s$ where $c_s$ is the
sound speed and $\ell_{\rm mix}$ is a free parameter we vary to study
the sensitivity of our results to turbulence. In the context of
accretion disk theory turbulent viscosity is typically parametrized in
terms of a dimensionless constant $\alpha$ linked to $\ell_{\rm mix}$
through the relation $\ell_{\rm mix} = \alpha\, c_s\, \Omega^{-1}$,
where $\Omega$ is the angular velocity of the fluid
\citep{shakura:1973a}.

Recently, \citet{kiuchi:2017zzg} performed very high resolution
\ac{GRMHD} simulations of a \ac{NS} merger with sufficiently high seed
magnetic fields $(10^{15}\, {\rm G})$ to be able to resolve the \ac{MRI}
in the merger remnant and reported averaged $\alpha$ values for
different rest-mass density shells. Combining their estimate of $\alpha$
with values of $c_s$ and $\Omega$ from our simulations we find values of
$\ell_{\rm mix} = 0{-}30\, {\rm m}$. These values are also consistent
with estimates based on dimensional considerations \citep{duez:2006qe,
radice:2017zta}. Here, we conservatively vary $\ell_{\rm mix}$ between
$0$ (no subgrid model) and $50\, {\rm m}$ (very efficient angular
momentum transport). On the other hand, we want to emphasize that these
estimates for $\ell_{\rm mix}$ have been derived for the post-merger
phase and under specific assumptions, such as strong initial magnetic
fields and equal masses, while the results presented here depend on the
effective viscosity present during the merger for unequal mass systems.
Moreover, we caution the reader that in reality $\ell_{\rm mix}$ is
likely to be time-dependent and non-constant. For these reasons, our
results should only be considered as qualitative until the relevant
$\ell_{\rm mix}$ can be estimated, and our approach validated with
\ac{GRMHD} simulations.
\acused{MHD}

We consider two binaries: an equal mass binary with component masses
$1.35\, M_\odot$ and $1.35\, M_\odot$, and an unequal mass binary with
component masses $1.4\, M_\odot$ and $1.2\, M_\odot$. Notably, this is
the first time an unequal mass \ac{NS} merger simulation has been
performed in GR including viscosity. We adopt the LS220 \ac{EOS}
\citep{lattimer:1991nc} that is based on a liquid droplet Skyrme model
and predicts a maximum mass of 2.06 $M_\odot$ and radius $R_{1.4}$ of
12.7 km for non-rotating cold \acp{NS}. Hence, the \ac{EOS} is
compatible with current astrophysical constraints, including the recent
LIGO/Virgo constraint on tidal deformability
\citep{theligoscientific:2017qsa, abbott:2018wiz, abbott:2018exr,
de:2018uhw}.

We evolve the initial data using the \texttt{WhiskyTHC} code
\citep{radice:2012cu, radice:2013hxh, radice:2013xpa, radice:2015nva}.
Neutrinos losses are modeled by a leakage scheme
\citep{galeazzi:2013mia, radice:2016dwd}, with a free-streaming
component evolved according to the M0 scheme introduced in
\citet{radice:2016dwd}. The M0 scheme models the re-absorption effects
and includes approximate gravitational and Doppler effects in a
computational efficient way. More technical details are reported in
\citet{radice:2018pdn}.

\section{Results}

The inclusion of viscosity affects the thermodynamical properties of the
merger remnants, their lifetimes, and the neutrino luminosities, as
discussed in \citet{radice:2017zta}. We focus here only on the dynamical
ejecta.

The total ejecta mass for the equal mass $(1.35 + 1.35)\, M_\odot$
binary does not show a systematic trend with viscosity: we find $0.19
\times 10^{-2}\, M_\odot$, $0.27 \times 10^{-2}\, M_\odot$, $0.20 \times
10^{-2}\, M_\odot$, and $0.20\times 10^{-2}\, M_\odot$ of dynamical
ejecta for the $\ell_{\rm mix} = 0$, $5\, {\rm m}$, $25\, {\rm m}$, and
$50\, {\rm m}$ simulations, respectively. These differences are at the
level expected from the stochastic nature of the mass ejection
\citep{bauswein:2013yna, radice:2018pdn}. The intensive properties of
the ejecta, \ie, electron fraction, entropy, or on their asymptotic
velocities are also rather insensitive to the inclusion of viscosity.
That said, we remark that with the inclusion of turbulent viscosity the
mass outflow rate does not drop completely to zero after the merger, as
is instead the case for the binaries simulated without the inclusion of
viscous angular momentum transport. Instead, the dynamical mass ejection
is immediately followed by the early phase of a secular viscous-driven
outflow \citep[\eg,][]{lee:2009a, fernandez:2013tya, metzger:2014ila,
siegel:2017nub, fujibayashi:2017puw, fernandez:2018kax}. Our simulations
do not extend sufficiently in time to allows us to study the secular
ejecta. This will be the object of our future work.

\begin{figure}
  \includegraphics[width=0.98\columnwidth]{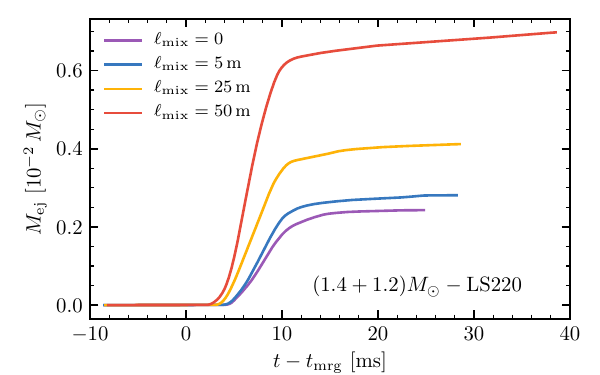}
  \caption{Mass of the dynamical ejecta for the $(1.4 + 1.2)\, M_\odot$
  binary simulated with different values of the viscous parameter
  $\ell_{\rm mix}$. We find that viscous effects can boost the dynamical
  ejecta mass by factors of a few.}
  \label{fig:total_ejecta_mass}
\end{figure}

The case of the unequal mass binary $(1.4 + 1.2)\, M_\odot$ is
qualitatively different and shows a new phenomenon.  We find that the
dynamical ejecta mass increases monotonically with the viscous parameter
$\ell_{\rm mix}$, as shown in Fig.~\ref{fig:total_ejecta_mass}. The
dynamical ejecta mass increases by almost a factor 3, from $0.24 \times
10^{-2} \, M_\odot$ to $0.7 \times 10^{-2}\, M_\odot$, as the mixing
length is increased from $0$ to $50\, {\rm m}$. The early phase of the
secular outflow rate after the merger also grows monotonically with the
viscosity. The amount of matter ejected between $t = 15$~ms and $t =
25$~ms, which we tentatively identify with the first part of the secular
ejecta, is $0.09\times 10^{-3}\, M_\odot$, $0.17\times 10^{-3}\,
M_\odot$, $0.20\times 10^{-3}\, M_\odot$, and $0.24\times 10^{-3}
M_\odot$ for the $\ell_{\rm mix} = 0$, $\ell_{\rm mix} = 5\ {\rm m}$,
$\ell_{\rm mix} = 25\, {\rm m}$, and $\ell_{\rm mix} = 50\, {\rm m}$
runs respectively. While the increase in the secular ejecta mass with
viscosity was expected \citep[\eg,][]{fujibayashi:2017puw}, the
significant growth in the dynamical ejecta mass with viscosity was not.

\begin{figure*}
  \begin{center}
    \includegraphics[width=1.96\columnwidth]{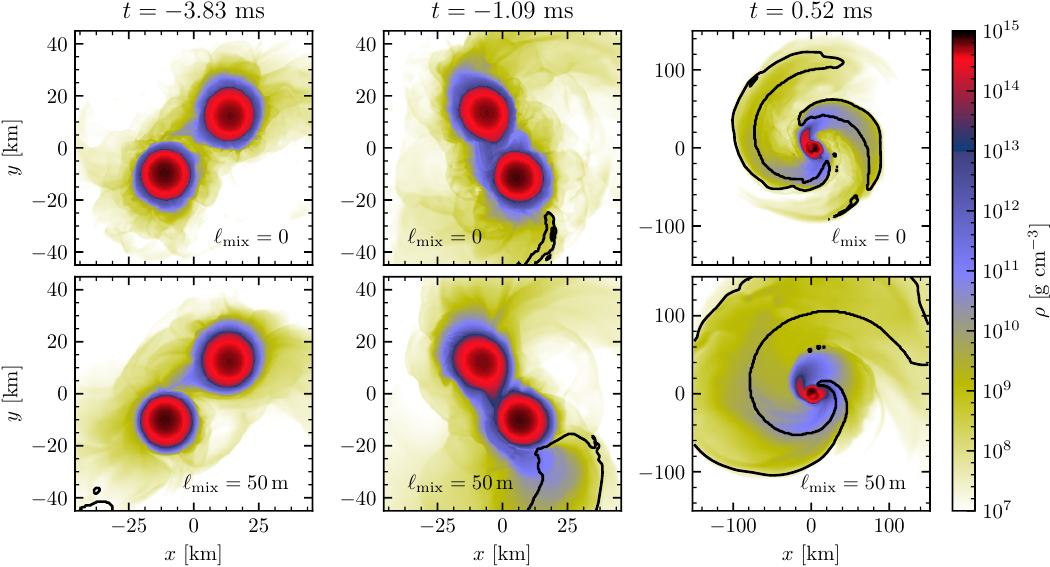}
    \caption{Rest mass density on the orbital plane for the $(1.4 +
    1.2)\, M_\odot$ binary simulated with different values of the
    viscous parameter $\ell_{\rm mix}$. The black contour denotes
    unbound material, \ie, with $-u_0 > 1$. Note that more material can
    become unbound at larger radii. Viscous heating of the tidal stream
    between the primary and the secondary stars can significantly
    enhance the mass loss from the binary.}
    \label{fig:ejecta_visc_2d}
  \end{center}
\end{figure*}

The origin of the ejecta mass enhancement can be understood from the
comparison of the density profiles on the orbital plane for the $\ell =
0$ and $\ell = 50\, {\rm m}$ runs show in Fig.~\ref{fig:ejecta_visc_2d}.
We observe the development of mass transfer between the secondary and
the primary \acp{NS} few milliseconds prior to the merger. In the
simulation with no viscosity the material outflowing from the secondary
settles on the primary \ac{NS}. The \acp{NS} are not corotating, so the
accreted material must form a layer on the surface of the primary
\ac{NS} characterized by a large velocity gradient. Viscous heating
heats this layer to high temperatures, up to ${\sim}20{-}30\, {\rm MeV}$
for $\ell_{\rm mix} = 50\, {\rm m}$. The resulting pressure gradient
drives the copious mass outflow. We stress that our fiducial binary has
a only moderate mass ratio of $q\simeq 0.85$. Mass transfers is expected
to increase for more asymmetric binaries, thus an even larger effect
could be expected for smaller $q$.

The viscous-dynamical ejecta have a broad distribution in $Y_e$ that is
very similar to the dynamical ejecta observed in simulations without
viscosity \citep{radice:2018pdn}. This is not too surprising given that
both the shocked and viscous-dynamical ejecta are composed of material
from the outer layers of the \acp{NS} that is pushed by hydrodynamical
forces operating from the interior of the forming merger remnant. In one
case the hydrodynamical push is generated by shocks launched after
merger, in the other by the pressure gradient induced by the viscous
heating of the primary \ac{NS}.

The distinctive features of the viscous-dynamical ejecta are their
larger mass and higher asymptotic velocities, the latter up to
${\sim}30\%$ larger than that for the shocked ejecta. Moreover, our
simulations shows that the amount of fast moving ejecta with asymptotic
velocities larger than $0.6\, c$ is also significantly increased with
the inclusion of viscosity. In the case of the $(1.4 + 1.2)\, M_\odot$
binary it grows monotonically from $10^{-8}\, M_\odot$ in the run with
no viscosity to $8.3\times 10^{-5}\, M_\odot$ for the run with
$\ell_{\rm mix} = 50\, {\rm m}$. A more modest, but still significant,
increase in the amount of fast moving ejecta is also observed for the
equal mass binary. In this case the amount of ejecta with asymptotic
velocities in excess of $0.6\, c$ grows from $10^{-7}\, M_\odot$ of the
simulation with no viscosity to $0.4\times 10^{-5}\, M_\odot$ of the
simulation with $\ell_{\rm mix} = 50\, {\rm m}$.

\section{Discussion}

\begin{figure*}
  \includegraphics[width=0.98\columnwidth]{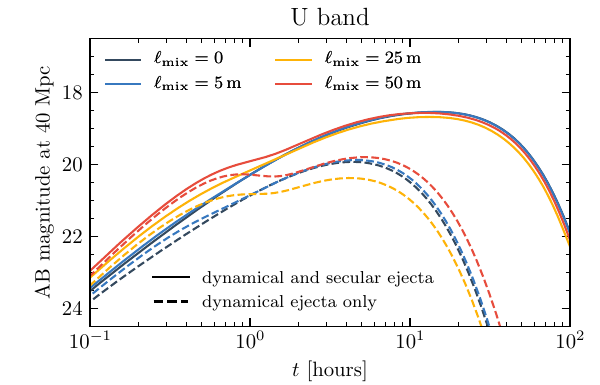}
  \hfill
  \includegraphics[width=0.98\columnwidth]{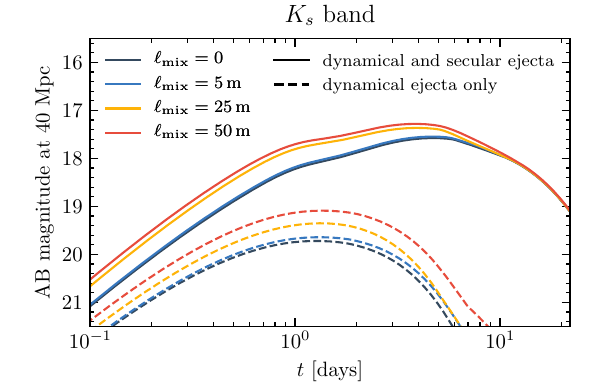}
  \caption{kN light curves for equatorial observers in the $U$ band
  (\emph{left panel}) and $K_s$ band (\emph{right panel}) for the $(1.4
  + 1.2)\, M_\odot$ binary simulated with different values of the mixing
  length $\ell_{\rm mix}$. We show light curves obtained accounting only
  for the dynamical ejecta (dashed lines), and light curves computed
  including contributions from both the dynamical and the secular ejecta
  (solid lines). The viscous-dynamical ejecta expands sufficiently
  rapidly so that free-neutron decay can occur and would leave a
  characteristic bump in the UV/optical light curves in the first hour.
  The viscous-dynamical ejecta also enhance the flux in the NIR bands at
  later times.}
  \label{fig:kilonova}
\end{figure*}

We compute synthetic \ac{kN} light curves using the semi-analytical
model of \citet{perego:2017wtu}. Composition, angular distribution, and
velocity of the dynamical ejecta are directly taken from the
simulations. We also include the contribution of the secular ejecta
which we assume to be composed of neutrino-driven wind entraining
$0.01\, M_\odot$ of high $Y_e$ material and of $0.05\, M_\odot$ of
intermediate $Y_e$ viscous outflows from the remnant's accretion disk.
Our model also includes the contribution of free-neutron decay following
\citet{metzger:2014yda} and \citet{metzger:2016pju}. In particular, we
assume that neutron capture is avoided for ejecta expanding with
velocities in excess of $0.6\, c$ \citep[see][Fig.~1]{metzger:2014yda}.
We refer to \citet{perego:2017wtu} and \citet{radice:2018pdn} for a
full account of all the inputs to our model.

When considering the $(1.4 + 1.2)\, M_\odot$ binary, we find that the
inclusion of viscosity results in a visible bump in the UV light curve
on a timescale of about one hour of the merger. This bump disappears if
we switch off the contributions from free neutrons to the emission of
the ejecta, so it would be a clear signature of the production of a fast
outflow. Overall we find that, because of the presence of free neutrons,
the \ac{kN} could be up to one magnitude brighter in the UV bands in the
first hour. This is a somewhat smaller enhancement than that reported by
\citet{metzger:2014yda}, who found that free neutrons could boost the
brightness of the \ac{kN} in the UV bands by up to 4 magnitudes. We
remark that the peak magnitude in $U$ band for our $\ell_{\rm mix} =
50\, {\rm m}$ model is consistent with the prediction of
\citet{metzger:2014yda} for the case with $10^{-4}\, M_\odot$ of free
neutrons, taking into account differences in assumed opacities, ejecta
masses, and expansion velocities. However, possibly because of the more
rapid expansion of our ejecta and the presence of low opacity material
at high latitude in our simulations, our synthetic \ac{kN} light curves
are brighter at early times compared with the baseline models of
\citet{metzger:2014yda}.

Due to the significant increase of the dynamical ejecta masses, the
viscous runs result in modest increases to the \ac{kN} light curves in
the \ac{NIR} bands by up to ${\sim}0.5$ magnitudes (see
Fig.~\ref{fig:kilonova}). The reason why the increase in the \ac{NIR}
flux is modest is that the \ac{kN} is actually dominated by radiation
emitted by the secular ejecta on the relevant timescales. This can be
seen by comparing the \ac{kN} light curves computed with the inclusion
of both secular and dynamical ejecta with those generated accounting
only for the latter. The increase in the amount of high photon opacity
dynamical ejecta due to viscosity is also inconsequential, because
this outflow component is not sufficiently massive and expands too
rapidly to significantly obscure the emission from the secular ejecta on
a timescale of more than few hours.

\begin{figure}
  \includegraphics[width=0.98\columnwidth]{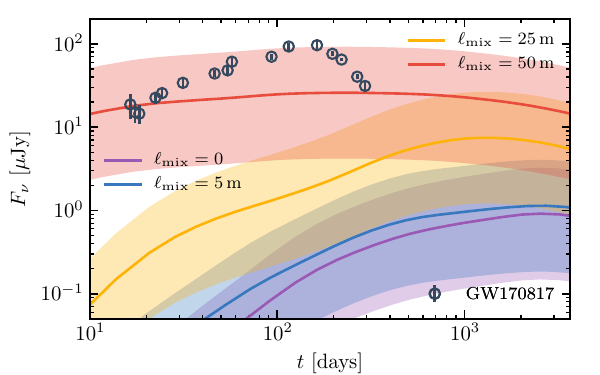}
  \caption{Radio light curves at 3~GHz for the $(1.4 + 1.2)\, M_\odot$
  binary simulated with different values of the mixing length $\ell_{\rm
  mix}$. The different bands correspond to different models were we vary
  the ISM density from $10^{-4}\, {\rm cm}^{-3}$ to $5 \times 10^{-3}\,
  {\rm cm}^{-3}$, a range compatible with current observational
  constraints \citep{mooley:2018qfh}. The solid lines correspond to the
  fiducial value $n = 10^{-3}\, {\rm cm}^{-3}$.  Also shown as open
  circles are the observed flux densities at 3~GHz of the afterglow in
  GW170817, which is presently dominated by the interaction between a
  relativistic jet and the ISM \citep{hallinan:2017woc, mooley:2017enz,
  mooley:2018qfh}. Here we assume the microphysics parameters to be
  $\epsilon_e=0.1$, $\epsilon_B=0.01$, and $p=2.16$. Viscous-dynamical
  ejecta could power very bright synchrotron remnants. The $\ell_{\rm
  mix} = 50\, {\rm m}$ simulation predicts a rebrightening of GW170817
  in the radio on a timescale of ${\sim}1$ year from the merger.}
  \label{fig:radio}
\end{figure}

We estimate the radio light curves from synchrotron emission generated
as the ejecta interacts with the \ac{ISM} using the semi-analytic model
of \citet{hotokezaka:2015eja}. The free parameters of this model are the
\ac{ISM} number density $n$, the efficiencies with which the internal
energy of the shock is converted into kinetic energy of non-thermal
electrons $\epsilon_e$ and magnetic energy of the amplified background
field $\epsilon_B$. The model assumes spherical symmetry and takes as
input the asymptotic velocity distribution of the ejecta.

Our results are shown in Fig.~\ref{fig:radio}. The radio light curves
are very sensitive to the asymptotic velocity distribution of the ejecta
\citep{hotokezaka:2018gmo} and, in particular, to the presence of a
high-velocity tail. They are also sensitive to the \ac{ISM} density. In
the case of GW170817 the latter is presently constrained to be
${\sim}10^{-4}-5 \times 10^{-3}\, {\rm cm}^{-3}$ \citep{mooley:2018qfh,
ghirlanda:2018uyx}. We find that the viscous-dynamical ejecta could
power radio remnants that are orders of magnitude brighter and peak at
earlier times than those predicted by simulations that do not include
viscosity. It might be possible to detect or exclude the presence of a
large amount of viscous-dynamical ejecta with continued radio
observations of GW170817 in the next few years. At the moment the
synchrotron emission from GW170817 is thought to be powered by the
interaction between a relativistic jet and the ISM
\citep{mooley:2018qfh, ghirlanda:2018uyx}, however the viscous-dynamical
ejecta might be detectable as a break in the decline of the radio data
occurring when the emission from the jet will have faded. For mergers
occurring in higher density environments, the radio flares generated by
the interaction between the viscous-dynamical ejecta and the \ac{ISM}
could be detected to distances of several tens to few hundreds Mpc.

\section{Conclusions}

We have shown that the effective viscosity possibly arising from
small-scale \ac{MHD} turbulence prior to merger can significantly boost
the amount of dynamical ejecta for unequal mass binaries. The resulting
outflows are neutron rich and have velocity distributions with large
mean values and extended tails. These viscous-dynamical ejecta would
power \acp{kN} showing signatures of free neutron decays on a timescale
of about one hour. Indeed, a fraction of the viscous-ejecta, up to
${\sim}10^{-4}\, M_\odot$, expands sufficiently rapidly for most
neutrons to avoid capture. The resulting heat from the beta-decays would
leave a detectable imprint on the \ac{kN} light curve at early times.
The viscous-dynamical ejecta would also contribute to the \ac{kN} signal
in the \ac{NIR} bands on a timescale of few days. Due to their large
kinetic energies, the outflows produced by this new mechanism could also
generate very bright radio remnants as they interact with the \ac{ISM}
on longer timescales of weeks to months. Accordingly, it might be
possible to constrain the mass and kinetic energy of the
viscous-dynamical ejecta for GW170817 with radio observations in the
coming months and years.

Whether or not the new ejection mechanism discussed here operates in
Nature depends on how large the effective viscosity of the mass exchange
flows are. At the moment, very rough estimates for the effective
viscosity exist only for the post-merger phase of \ac{NS} mergers, while
the mechanism discussed here depends on the effective viscosity of the
flow at the time of merger. Understanding this will require very
high-resolution and/or local \ac{GRMHD} simulations. This will be the
object of future work.

\begin{acknowledgments}
It is a pleasure to acknowledge A.~Burrows for the many stimulating
discussions, and B.~Metzger for comments on an earlier draft of the
manuscript.
DR gratefully acknowledges support from a Frank and Peggy Taplin
Membership at the Institute for Advanced Study and the
Max-Planck/Princeton Center (MPPC) for Plasma Physics (NSF PHY-1523261).
AP acknowledges support from the INFN initiative "High Performance data
Network" funded by CIPE.  DR and AP acknowledge support from the
Institute for Nuclear Theory (17-2b program) and from the
Theory Alliance-Facility for Rare Isotope Beams
(Topical Program: FRIB and the GW170817 kilonova).
AP thanks the Institute
for Advanced Study for its hospitality and support. SB acknowledge
support by the EU H2020 under ERC Starting Grant, no.~BinGraSp-714626.
SAF acknowledges support from the United States Department of Energy
through the Computational Science Graduate Fellowship, grant number
DE-SC0019323. LFR
acknowledges support from U.S. Department of Energy through the award
number DE-SC0017955.
The simulations were performed on BlueWaters, Bridges, Comet, and
Stampede, and were enabled by the NSF PRAC program (ACI-1440083 and
AWD-1811236) and the NSF XSEDE program (TG-PHY160025). The analysis
employed computational resources provided by both the TIGRESS high
performance computer center at Princeton University, which is jointly
supported by the Princeton Institute for Computational Science and
Engineering (PICSciE) and the Princeton University Office of Information
Technology, and the Institute for Cyber-Enabled Research, which is
supported by Michigan State University.
\end{acknowledgments}

\end{document}